# PATTERNS OF URBAN FOOT TRAFFIC DYNAMICS

GREGORY DOBLER[†,1234], JORDAN VANI[4], TRANG TRAN LINH DAM[4]

*October 4, 2019*

ABSTRACT

Using publicly available traffic camera data in New York City, we quantify time-dependent patterns in aggregate pedestrian foot traffic. These patterns exhibit repeatable diurnal behaviors that differ for weekdays and weekends but are broadly consistent across neighborhoods in the borough of Manhattan. Weekday patterns contain a characteristic 3-peak structure with increased foot traffic around 9:00am, 12:00-1:00pm, and 5:00pm aligned with the "9-to-5" work day in which pedestrians are on the street during their morning commute, during lunch hour, and then during their evening commute. Weekend days do not show a peaked structure, but rather increase steadily until sunset. Our study period of June 28, 2017 to September 11, 2017 contains two holidays, the 4th of July and Labor Day, and their foot traffic patterns are quantitatively similar to weekend days despite the fact that they fell on weekdays. Projecting all days in our study period onto the weekday/weekend phase space (by regressing against the average weekday and weekend day) we find that Friday foot traffic can be represented as a mixture of both the 3-peak weekday structure and non-peaked weekend structure. We also show that anomalies in the foot traffic patterns can be used for detection of events and network-level disruptions. Finally, we show that clustering of foot traffic time series generates associations between cameras that are spatially aligned with Manhattan neighborhood boundaries indicating that foot traffic dynamics encode information about neighborhood character.

## 1. INTRODUCTION

With the increasing urbanization of global populations, the understanding of cities as dynamical systems is becoming increasingly important for future planning and sustainable growth of urban environments. And while the impact that built infrastructures have on resource consumption and the local to global environment is dependent on technological advancement of engineered systems, it is the dynamical behavioral patterns of the underlying urban populations that serve as the primary driver for city functioning. These patterns of human dynamics in cities show characteristic macro-/micro-scale behavior [1] and are seen across many sectors including energy consumption [2] and transportation [3]. The macro-scale, aggregate behavior in particular has been shown to exhibit diurnal, weekly, monthly, and seasonal patterns that manifest in a variety of ways including lighting variability [4], taxi pick-ups and drop-offs [5], social media check-ins [6], and public WiFi pings [7]. The latter represents an example of tracers of pedestrian foot traffic and human mobility [8-10], a critical aspect of understanding pedestrian use of – and flow through – public spaces [11-17].

[†] gdobler@udel.edu
[1] Biden School of Public Policy and Administration, University of Delaware
[2] Department of Physics and Astronomy, University of Delaware
[3] Data Science Institute, University of Delaware
[4] Center for Urban Science and Progress, New York University

Human mobility in cities has been studied in a variety of contexts that includes origin-destination mapping [18,19], transportation infrastructure use [20], and pedestrian behavior [21]. This pedestrian behavior, in turn, informs key operational and quality of life indicators in cities including public safety [22,23] and public health [24,25], and serves as an input for models of urban planning relating to the lived experience of cities [26-29]. Methods for assessing pedestrian foot traffic in cities are numerous and include the aforementioned WiFi pings [7,30,31] as well as Bluetooth activity [30,31], thermal and laser sensing [32], and video-based methods [33-36]. While each of these has advantages and disadvantages with respect to accuracy and bias, video/imaging methods in which humans are detected in single or sequences of images have the most potential for reasonably accurate counting and trajectory determination [37-39] with a minimum number of deployed sensors. The disadvantages of imaging-based methods include significantly increased computational complexity to detect a pedestrian; substantially higher data rates; and privacy, ethical, and legal considerations of deployed imaging systems in public spaces [40,41] that can potentially identify individuals via facial [42] or gait [43] recognition[5]. Nevertheless, with appropriately trained and constructed pedestrian detectors as well as appropriately scoped privacy protections [47,48], aggregation of pedestrian detection in street-level imaging has the potential to critically inform a variety of urban disciplines including planning, transportation, and emergency management.

In this paper, we will demonstrate that using a trained classifier on streams of images from a network of traffic cameras in New York City (NYC) can be used to measure characteristic dynamical patterns of foot traffic in dense urban environments, providing city-scale information about public space usage. In §2 we describe the data and methodology used to implement the pedestrian detector, in §3 we present the observed aggregate patterns of activity and several anomalous cases, and in §4 we conclude with a discussion of the broader implications and ancillary urban science that is accessible from these data.

## 2. DATA AND METHODOLOGY

### 2.1 NYC Department of Transportation Traffic Cameras

The Department of Transportation in the city of New York (NYCDOT) maintains a network of traffic cameras, that are spread throughout the city. Although all five boroughs have at least one traffic camera, the primary concentration is on the island of Manhattan. Figure 1 shows the spatial locations of the camera network which consists of ~700 cameras for which data is publicly available [49]. Data from the traffic cameras is streamed at a rate of roughly 1 frame per second with a pixel resolution of 240x352. While this frame rate is largely insufficient for tracking or trajectory analysis (e.g., [50-52], but see [53]), it does allow for the instantaneous counting of vehicles and pedestrians within the field of view. Several examples of traffic cameras that are suitable for pedestrian counting as well as ones that are not are shown in Figure 1.

To generate the results presented in §3, we created a pipeline for data scraping and analysis that consisted of downloading a single image from a camera, counting pedestrians in that image (see §2.2), discarding the image, and moving on to the next camera. This pipeline was run continuously with a characteristic return time to the first camera of ~80s, implying that our sampling rate per camera is typically > 1 minute and that the counts are offset in time from one camera to the next. The data presented in this paper was collected and analyzed between June 28, 2017 and September 11, 2017. Using publicly available weather data, we exclude days that have >0.05 inches of precipitation. In total, our sample contains 31 weekdays (two of which are the 4th of July and Labor Day holidays) and 9 weekend days.

---

[5] Though WiFi and Bluetooth sensing that capture mac addresses is now also collecting Personally Identifiable Information [44] under recently enacted Privacy Laws in California, USA [45] and the European Union [46].

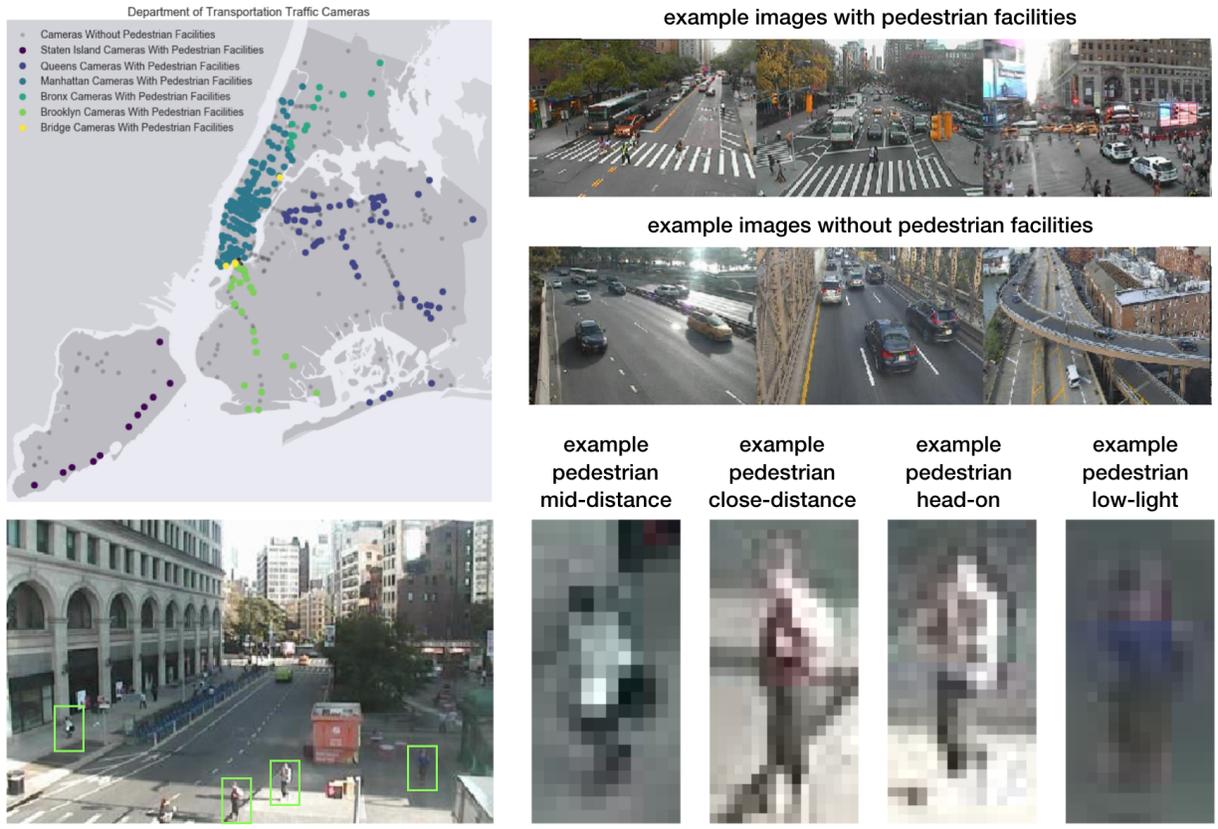

**Figure 1** – *Top left:* the geographic locations of the NYC Department of Transportation public traffic cameras. The cameras have coverage throughout all 5 boroughs but are concentrated in Manhattan. *Top right:* example images showing cameras both with and without views of pedestrian walkways. The former are the focus of analysis for this paper. *Bottom row:* an example image and zoomed in views of the pedestrians in that image demonstrating that the resolution of the traffic camera feeds is extremely low and does not contain personally identifiable information.

Details of the accuracy of the pedestrian counts is given below, however we stress that privacy protections are built into several key aspects of this work. First, the spatial resolution of the publicly available imaging is extremely low and does not allow for the extraction of personally identifiable information (PII) [44] including facial or gait recognition, nor is the resolution sufficient for license plate identification. A zoomed-in example of several images of pedestrians is shown in Figure 1. Second, our pedestrian detection model does not use full resolution feature maps as described in [54] but rather has multiple instances of down-sampling in the detection pipeline. In addition, our detector does not rely on features that contain PII or that can be used to match one detection to another. Third, we do not store all downloaded data while running our pedestrian counting pipeline. Roughly every 1 out of 1000 images is stored for consistency and accuracy checks, the rest are immediately deleted following application of the pedestrian detector. Lastly, we are presenting results from data collected >2 years in the past, eliminating the possibility that an individual's recent location is in the data set. We note also that, all of the results presented in this paper are aggregated in time.

*2.2 Human Detection in Video*

The field of pedestrian detection in both still images and video sequences of frames has both a long established history as well as breadth in methodology. A complete overview of that history is beyond the scope of this text (see [36] and references therein), but efforts to detect humans in video frames can be broadly separated into either hand engineered feature extraction plus a learned classifier or a deep convolutional network (CNN) model for automatic feature identification and classification. The most successful hand engineered feature-based models use the Histogram of Oriented Gradients (HOG) [55] with a robust classifier [56], however most modern pedestrian detection models are based on features learned through CNNs.

A significant advancement in the field of object detection, localization, and classification came through the development of Region-CNN (R-CNN) [57] which hand engineered features to identify regions of interest, merged regions according to their overlap, and finally implemented a CNN classifier on the regions identified. This model was subsequently improved in two iterations: Fast R-CNN [58] and Faster R-CNN [59]. The latter implements a full end-to-end neural network for region identification and classification, with training data that consists of bounding-boxes and labels for objects within images. This Faster R-CNN model was used in the data analysis pipeline for this paper as it was the state-of-the-art at the time that our data processing pipeline was deployed. Since the development of Faster R-CNN in 2015, there have been many further developments of models both similar [60] and dissimilar to Faster R-CNN, including those which use a combination of motion and still features [39,61], parts-based detectors [62], high dimensional features [63] and feature cascades [64], and recent models that use Faster R-CNN to extract high resolution features but replace the CNN classifier with tree-based methods [54].

*2.3 Model Training and Performance*

Our initial training/testing set was created from 3,918 daytime images that were pulled from 17 cameras on April 30th, 2016 and June 19th, 2016. The images were labelled by hand for positive pedestrian examples and negative examples using bounding boxes (BBs) with a constant aspect ratio of 3:4. These labels were not exhaustive (i.e., not all pedestrians were labelled). Across the 3,918 images, 16,022 positive and 41,449 negative examples were labelled yielding approximately a 2:5 (pos:neg) ratio. The enhanced number of negative examples was necessary for training in order to capture the complexity of the urban background. A Faster R-CNN model was trained on this data using the VGG16 network structure, a learning rate of 0.0005, a Region Proposal Network (RPN) batch size of 256, an RPN positive overlap of 0.7, and a minimum RPN size of 2x2 (smaller than typical given our very low resolution images). The network was trained for a total of 90,000 iterations on a GeForce GTX 1080 Ti GPU.

Model performance was assessed using a test set from the original labeled data by feeding the test set through the network and comparing centroids of labeled BBs with the BBs produced by the model through a pairwise elimination process of labels (points) to detections (boxes). Negative label centroids found outside all detection BBs were counted as a true negatives, negative label centroids within a detection BB were counted as false positives, positive label centroids found within a detection BB were counted as true positives, and positive label centroids not found within any detection BB were counted as false negatives. This process was performed iteratively through labeled centroids and detection BBs were removed when associated with a given true positive. A separate validation set was created using 286 randomly selected daytime and nighttime images from 3 distinct cameras. This validation set was then exhaustively labeled for pedestrians and a similar process of pairwise elimination was used to count true and false positives and negatives. The final precision and recall for our model was found from this validation set to be 92% ± 1.7% (precision) and 43 ± 1.9% (recall) where the uncertainties represent a 95% confidence interval found via 14-fold bootstrap resampling without replacement.

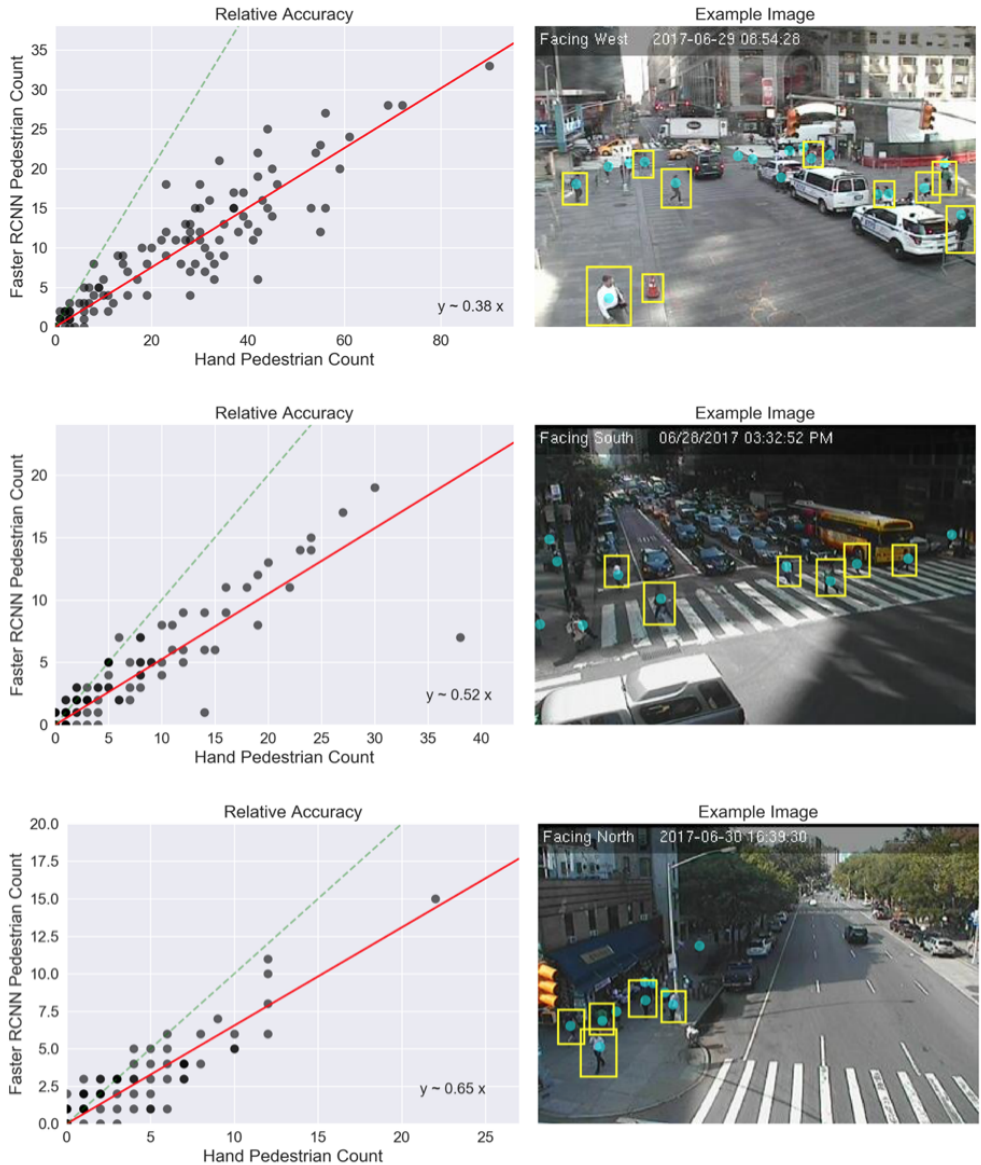

**Figure 2** – A comparison between the hand-counted number of pedestrians and the output of the detector (*left* panels) for several different cameras (*right* panels) at different times. Boosting the precision of our models to ensure few false positives results in a systematic undercounting. However, the figure shows that the true number of pedestrians scales linearly with the detected number for a given camera, thus allowing us to compare relative amplitudes of the number of pedestrians as a function of time as shown in Figure 3.

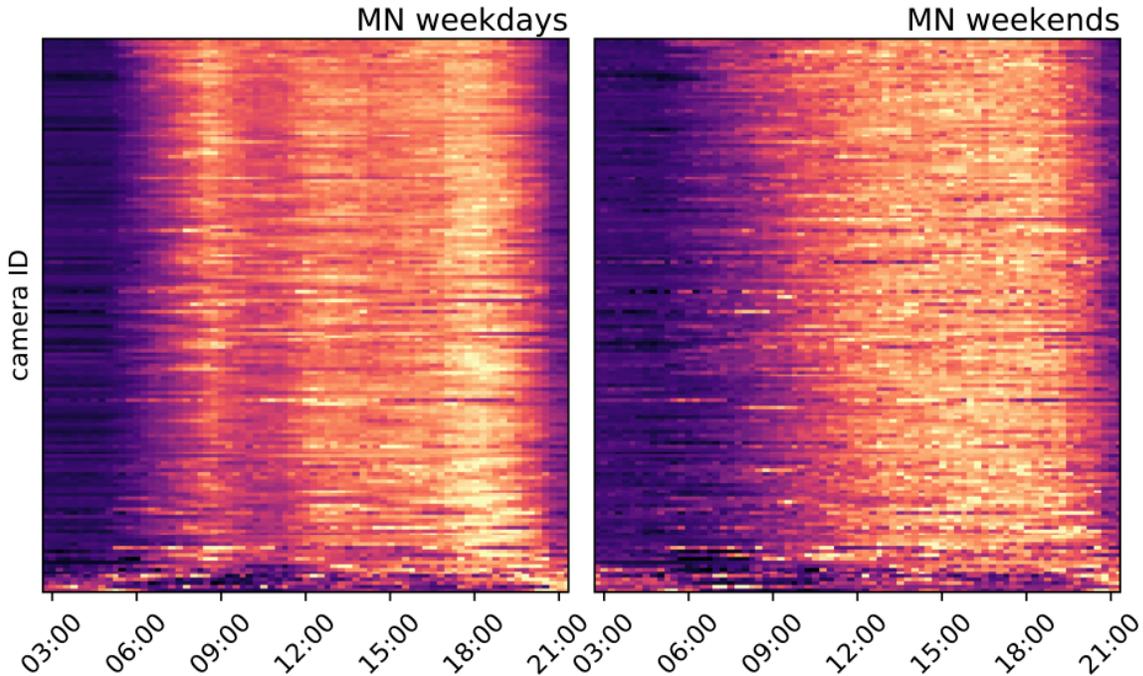

**Figure 3** – The standardized number of pedestrians detected in a given camera (rows) in Manhattan as a function of time of day from 3:00am to 9:00pm, averaged over all weekdays (*left*) and weekends (*right*) in our study period of June 28, 2017 to September 11, 2017. Weekday foot traffic dynamics display a 3-peak behavior that is strongly aligned with the "9-to-5" workday in which the peaks correspond to commuting to work, exiting buildings during the lunch hour, then commuting from work. The weekend dynamics do not show a peaked structure, but rather a steady increase of pedestrian counts until night time.

While the precision of the model is reasonably high, the relatively low recall indicates that, in a given image, we are undercounting the number of pedestrians. However, as we will show in §3, our primary interest is in measuring trends and dynamics and so absolute numbers are less important than *relative* amplitudes over time. In that case, it suffices to show that the true number of pedestrians in the field of view of an image from a given camera scales linearly with the detected number of pedestrians in that camera's image. Figure 2 demonstrates that this is indeed the case. For the three cameras shown, we counted the number of pedestrians in the field of view and find that this number scales linearly with the detected number, though the coefficient of correlation varies from camera to camera due to both scene and viewing angle variability.

## 3. RESULTS

Throughout the rest of this paper, we will restrict our results to the borough of Manhattan due to both the high density of NYCDOT cameras in that borough and the relatively large number of pedestrians at a given camera location. The latter ensures that we have sufficient statistics to identify the patterns presented below. We also bin the number of detections to 15 minute intervals for each camera to reduce noise prior to standardization of the time series or averaging across days or cameras.

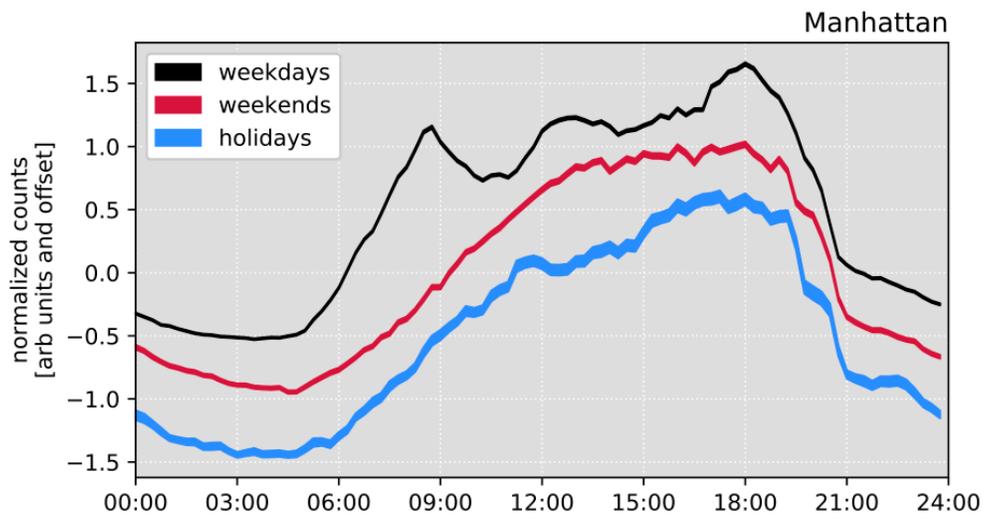

**Figure 4** – The same as Figure 3, but averaged over cameras in Manhattan with the width of the line corresponding to the error on the mean at a given time of day. Our study period included two holidays that fell on a weekday, the 4th of July and Labor Day. Averaging over Manhattan cameras for those days shows that holiday foot traffic dynamics display weekend as opposed to weekday behavior.

*3.1 Foot Traffic Dynamics*

Figure 3 shows the diurnal variability of pedestrian counts for each camera in Manhattan averaged over both weekdays and weekends during our sample period. Each row represents a given camera location and the total counts have been standardized after averaging across days. Two patterns in the foot traffic dynamics are clearly apparent: weekdays exhibit a characteristic 3-peak structure while weekends show a steady increase throughout the day. Our interpretation of the 3-peak weekday behavior is that the cameras are detecting morning rush as pedestrians walk through the streets to work, a "lunch hour" bump during which pedestrians temporarily exit buildings out into the streets, and an afternoon rush hour as pedestrians leave work for home. That this 3-peak structure is so characteristic to weekdays compared to weekends would indicate that it is tightly related to work schedules.

To test this hypothesis, we show the average weekday and weekend across all cameras in Manhattan in Figure 4 as well as the average *holiday*. Over our sample period, there were two such holidays, July 4th and Labor Day, which fell on a Tuesday and Monday respectively in 2017. These holidays (in which many businesses are closed) do not exhibit the 3-peak structure and align much more closely with the no-peak weekend structure, providing evidence that the 3-peak structure is closely tied to workforce behavior [3,6,7].

However, not all weekdays are identical. As we show in Figure 5, averaging over cameras in Manhattan for each day of the week shows that the transition from 3-peak to no-peak patterns is gradual as the week progresses. Monday through Wednesday shows clear 3-peak dynamical behavior, and the relative height of the middle bump becomes less prominent in the transition towards the weekend as the structure begins to break down by Friday. It is important to note that the majority of our sample period is during summer days and so it may be that many workers have curtailed Friday schedules, or that there is increased tourist activity on Fridays, which may lead to Fridays exhibiting characteristically "mixed" weekday/weekend behavior. This is underscored in Figure 6 where we have regressed each full day of our sample against the average weekday (Friday inclusive) and weekend day:

$$D_i = k_i K + e_i E \qquad (1)$$

where $D_i$ is the time series for a given day $i$, and $K = \sum_i K_i / N_K$ and $E = \sum_j E_j / N_E$ are the average weekday and weekend days respectively. That is, each time series has an associated projection onto the weekday/weekend phase space $(k_i, e_i)$. Figure 6 shows that, as the week progresses, the time series of urban foot traffic moves through this phase space with Fridays representing a mixture of weekday and weekend behavior. As described above, the two holidays (4th of July and Labor Day) have high "weekend" coefficients and low "weekday" coefficients despite both falling on a weekday. Interestingly, July 3rd, which fell on a Monday, shows a mix of weekday/weekend behavior indicating that many urban inhabitants did not go to work on the Monday before the 4th of July and/or that there was increased tourist activity on that day.

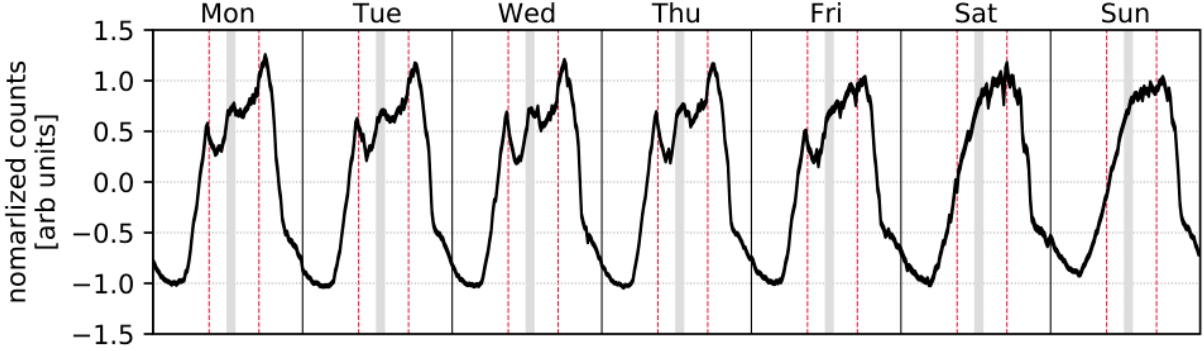

**Figure 5** – The standardized pedestrian foot traffic averaged over cameras in Manhattan during our sample period as a function of day of the week. Red vertical lines are shown at 9:00am and 5:00pm and the gray band covers the 12:00pm-1:00pm lunch hour. The 3-peak weekday pattern is strongest early in the week, but the peaks decrease in amplitude towards the end of the week as the pattern transitions to weekend behavior. Friday represents a mixture of weekday and weekend foot traffic patterns.

Lastly, Figure 6 also demonstrates the potential for $D_i$ to encode neighborhood character. Through straightforward K-means clustering of the weekday foot traffic time series (with 4 clusters that show different relative heights and positioning of the 3 peaks), we find that groupings of Manhattan foot traffic dynamics characteristically trace zonal boundaries in the borough: the Financial District (cluster 2), Greenwich Village and the Upper East/West Side (cluster 3), Midtown (clusters 0 and 1), and Harlem (cluster 0). This is perhaps not too surprising given the likelihood that the structures shown in Figures 3-5 relate to workforce and tourist behavior which are strong classifiers of these Manhattan neighborhoods.

*3.2 Anomaly and Event Identification*

The dynamical patterns of foot traffic described above have sufficient structure to identify anomalous outliers and events specific to a given camera location. In Figure 7 we show each full weekday (weekend day) in our sample period with the average weekday (weekend day) removed. All time series are averaged over Manhattan cameras. This residual shows strong anomalies on the holidays of the 4th of July and Labor Day, as well as a network-wide camera outage on July 27th between 2:15pm and 2:45pm. In addition, the figure demonstrates that our detection accuracy is affected by illumination with clear patterns of under-detection that correlate with sunset (gray line in the figure), though it is important to note that the earliest time that sunset may be affecting our determinations is 7:00pm (towards the end of our study period), significantly later in the day than the observed 3-peak structure, ensuring that the third peak in the weekday behavior is robust to variation in sunset time.

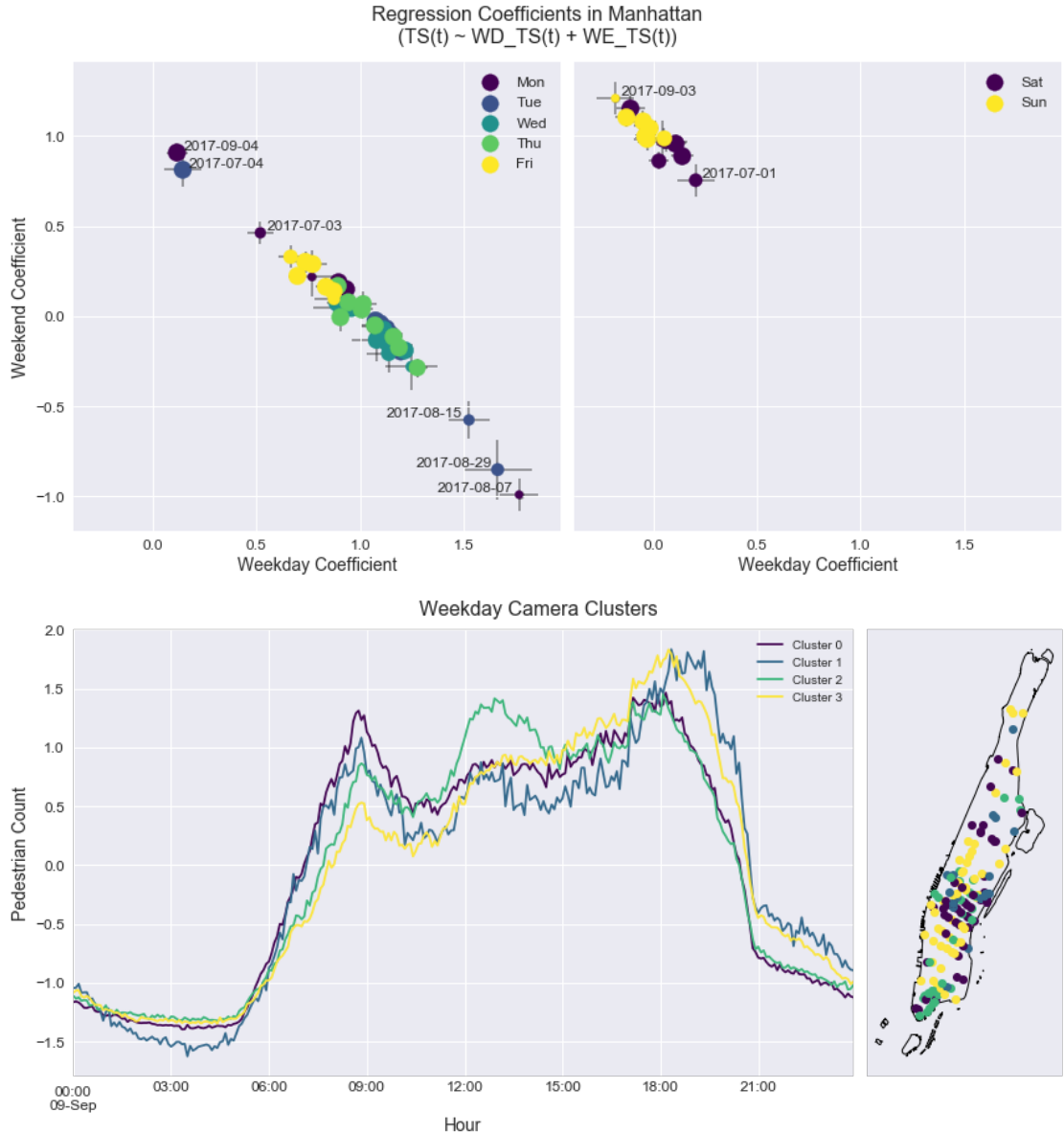

**Figure 6** – *Top:* a projection of each day in our study period onto weekday/weekend phase space in which each day is regressed against the average weekday and weekend foot traffic pattern. The size of each point is inversely proportional to the amount of rainfall on that day. The trajectory of weekdays through phase space demonstrates that Fridays represent a mixture of weekday and weekend behavior while holidays (the 4th of July and Labor Day) are well represented by weekend behavior. *Bottom:* Unsupervised clustering of the weekday time series for each camera into 4 clusters results in associations that spatially align with neighborhood boundaries indicating that foot traffic dynamics encode neighborhood character.

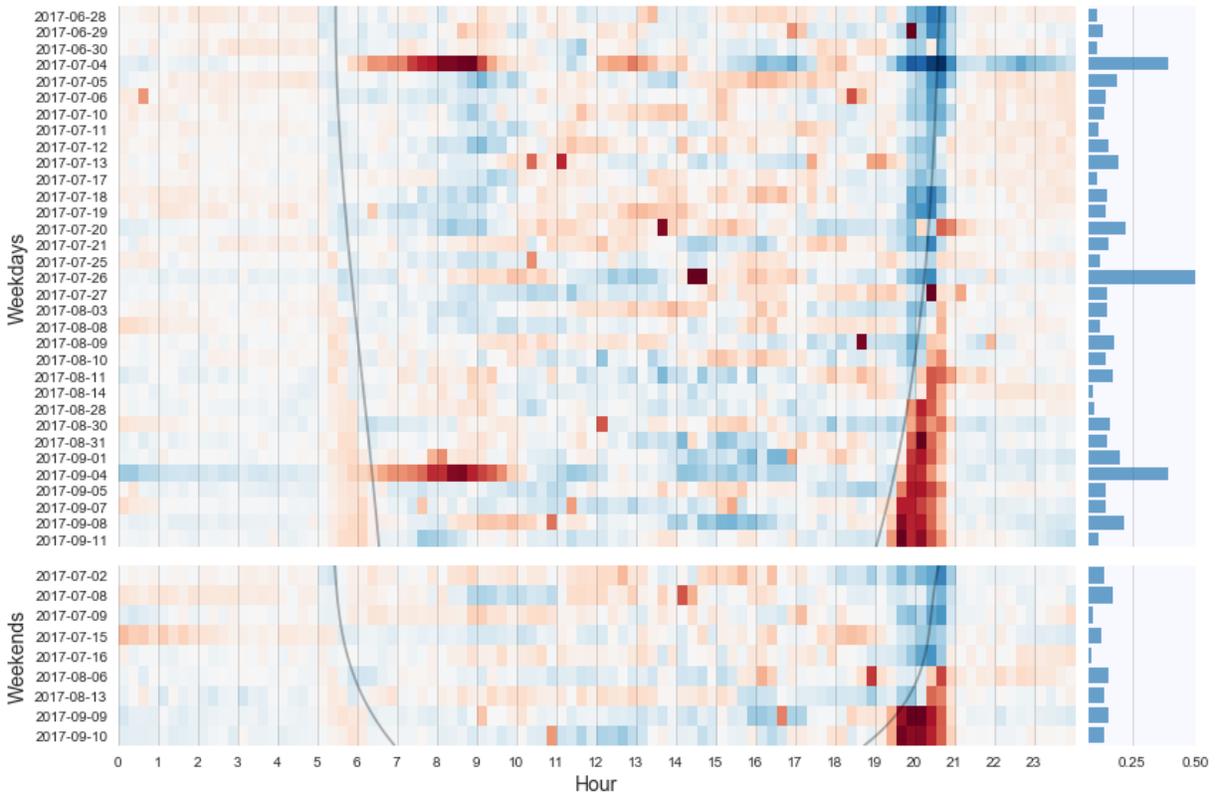

**Figure 7** – The residual of each day in our sample period (averaged over cameras in Manhattan) minus the average weekday (*top*) or weekend (*bottom*). The sum of squared residuals for each day is shown in the right panel. The gray lines represent sunrise/sunset time and these residuals show that our pedestrian detector is affected by lighting after sunset, however this is well after the third peak in the weekdays shown in Figures 3-5. These residuals do show the anomalous behavior of holidays as well as a system-wide camera outage for ~30 min on July 26th, 2017.

Beyond network-wide anomalies, for densely populated areas, we have sufficient signal-to-noise to detect anomalous (aggregate) behavioral dynamics at individual locations. We show two examples in Figure 8. In the first, the relative number of detected pedestrians between the hours of 3:00pm and 9:00pm is larger on the 4th of July (a holiday) than on weekend days for a camera located at 3rd Avenue and 42nd Street. We attribute this difference to the close vicinity of a viewpoint for the 4th of July fireworks near the East River Water Front, suggesting that this methodology can be used for onset of crowd detection. The second example is from a location near the Staten Island station in lower Manhattan on a weekday. The time series shows the characteristic 3-peak behavior, but superimposed on that are sharp spikes in the pedestrian counts at quasi-periodic intervals. The figure shows that these sharp features are tightly aligned with the ferry schedule indicating the utility of this method as a means of indirectly estimating transportation ridership.

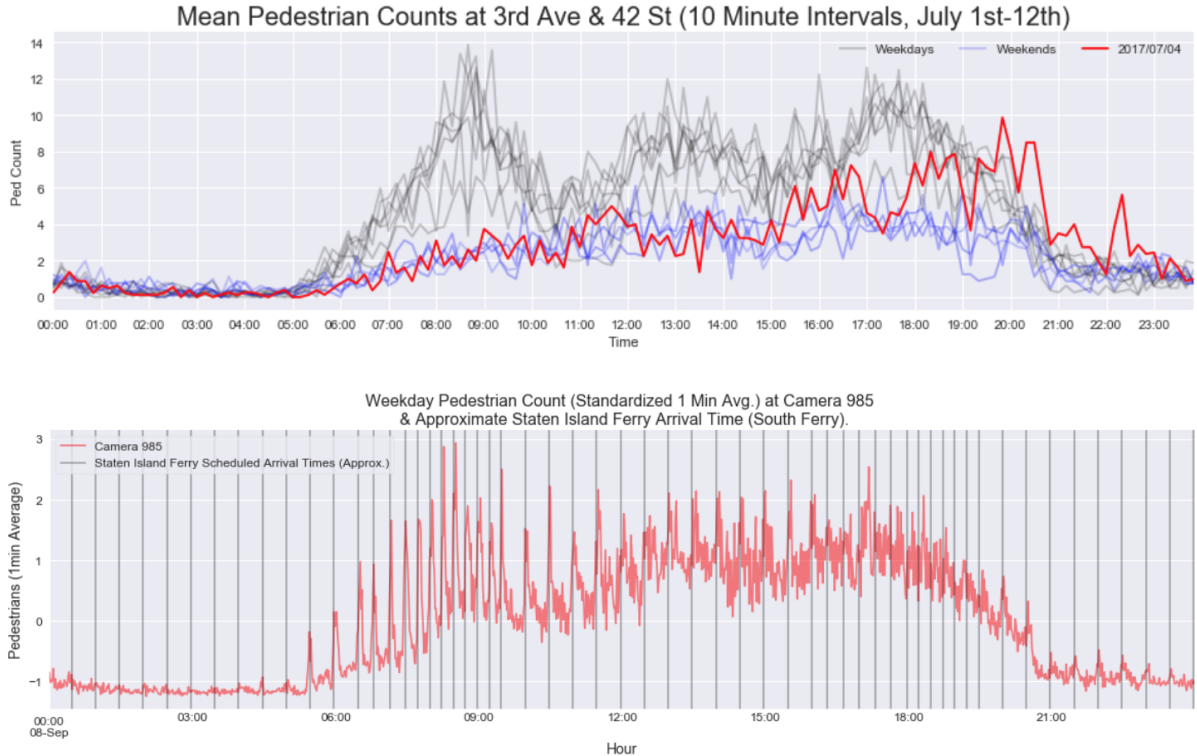

**Figure 8** – Examples of event detection in the time series of foot traffic for individual traffic cameras. *Top:* The weekdays (gray), weekend days (purple), and 4th of July (red) for the camera located at 3rd Avenue and 42 Street in Manhattan. As shown in Figure 7, the 4th of July is characteristically more similar to weekend than weekday behavior, but for this camera we find that the amount of pedestrian foot traffic increases significantly between the hours of 4:00pm and 8:00pm. We attribute this to the fact that this camera is near a viewpoint for the NYC 4th of July fireworks display and so this increase represents a gathering crowd of spectators. *Bottom:* For a camera near the exit of the Staten Island Ferry station in lower Manhattan, shows not only the 3-peak weekday pattern, but also sharp spikes in the number of pedestrians on the street that are tightly aligned with the ferry schedule (vertical gray lines), suggesting that this methodology can be used to indirectly estimate public transportation ridership.

4. CONCLUSIONS

We have presented a direct measurement of patterns in the dynamics of foot traffic in dense urban areas via the application of object detection methods to a network of low resolution traffic cameras in New York City. For a given camera, we find that the counts returned by our detector scale linearly with the actual number of pedestrians in the field of view allowing us to determine scaled trends in foot traffic activity for a given location. We have identified distinctly different foot traffic behavior on weekends vs weekdays when averaging across all cameras in the borough of Manhattan. Weekdays characteristically exhibiting a 3-peak structure, the peaks of which align with the onset of the "9-to-5" workday, the lunch hour, and the end of the workday. Weekend behavior typically does not show any such peaks but rather, foot traffic gradually increases throughout the day. Foot traffic behavior on holidays that fall on weekdays (the 4th of July and Labor Day in our June 28, 2017 to September 11, 2017 study period) traces weekend as opposed to weekday behavior. When aggregating across Manhattan and across the study period, we find that the 3-peak structure correlated with the work day is strongest early in the week, but Fridays display a mixture of weekday and weekend behavior. This is further demonstrated by tracing the trajectory of all study days through weekday/weekend phase space in which each day is projected onto dimensions representing the characteristic weekday and weekend temporal behavior. Finally, we have shown that there is evidence that unsupervised clustering of the foot traffic time series results in clusters that are *spatially* distinct and trace the boundaries of neighborhoods in Manhattan, indicating that foot traffic dynamics encode elements of neighborhood character.

Deviations from the established patterns of activity can be used to detect both system-wide anomalies and individual events. Examples of the former include weekday holidays that are clear outliers in weekend/weekday phase space as well as system-wide camera outages that present as simultaneous non-detection of pedestrians across all cameras. We have shown examples of event detection that include transitions to crowd behavior for fireworks displays on the 4th of July and recurrent bursts of detections near public transportation (PT) points that temporally align with PT schedules, corresponding to riders entering/exiting PT access points.

It is important to note that we have taken numerous steps to ensure the privacy of pedestrians in public within the field of view of the traffic cameras. First, the imaging data that was analyzed for this work was of sufficiently low resolution that it contained no personally identifiable information (PII) such as license plates or faces. Second, the analysis was done in near-real time with data discarded immediately after pedestrian counts were performed so that continuous data streams have not been stored. Third, the pedestrian detections are completely anonymous and there is no element of the analysis that attempts to either match a detection across multiple cameras or across the same camera in time (e.g., tracking was not performed). Finally, we are presenting the results of this study with a two year delay.

The measurements of city-scale foot traffic dynamics presented here have potential impacts for a wide range of urban operations including emergency management, situational awareness, transportation efficiency, planning of built environment interventions, and connectedness of urban communities. The scaling and deviation of these patterns across the city and correlations of those with characteristics of land use and neighborhood characteristics will be the subject of future work.

**Acknowledgements**
We thank Denis Khryashchev, Priya Kohker, Alec McLean, Richard Nam, and Bilguun Turboli for assistance generating the training set. GD, JV, and TTLD were supported by a Complex Systems Scholar Award from the James S. McDonnell Foundation.